\documentclass[a4paper,twocolumn,11pt]{quantumarticle2}
\pdfoutput=1
\usepackage[utf8]{inputenc}
\usepackage[english]{babel}
\usepackage[T1]{fontenc}
\usepackage{pstricks}
\usepackage{graphicx}
\usepackage{xcolor}
\usepackage{amsmath}
\usepackage{amssymb}
\usepackage{hyperref}
\usepackage{physics}
\usepackage{mathtools}
\usepackage{lipsum}
\newtheorem{theorem}{Theorem}
\newtheorem{example}{Example}
\newtheorem{definition}{Definition}
\newcommand{\qedwhite}{\hfill \ensuremath{\Box}}
\begin{document}
\title{\begin{minipage}{1.0\linewidth} Constructions and decoding procedures for quantum CSS codes \end{minipage}}
\author{Yannick Saouter}
\affiliation{Laboratoire Lab-STICC (UMR CNRS 6285), Technop\^ole
  Brest-Iroise,CS 83818 29238 Brest C\'edex 3, France.}
\orcid{0000-0001-7980-9168}
\email{yannick.saouter@imt-atlantique.fr}
\author{Massinissa Zenia}
\email{massinissa.zenia@univ-brest.fr}
\affiliation{Universit\'e de Bretagne Occidentale, Lab-STICC, CNRS,
  UMR 6285, 6 avenue Le Gorgeu, 29200 Brest, France.}
\author{Gilles Burel}
\affiliation{Universit\'e de Bretagne Occidentale, Lab-STICC, CNRS,
  UMR 6285, 6 avenue Le Gorgeu, 29200 Brest, France.}
\orcid{0000-0002-1427-4577}
\email{Gilles.Burel@univ-brest.fr}
\maketitle
\begin{abstract}
This article presents new constructions of quantum error correcting
Calderbank-Shor-Steane (CSS for short) codes. These codes are mainly
obtained by Sloane's classical combinations of linear codes applied
here to the case of self-orthogonal linear codes.
A new algebraic decoding technique
is also introduced. This technique is exemplified on CSS codes
obtained from BCH, Reed-Muller and projective geometry codes.
\end{abstract}
\section{Introduction}
Quantum technologies form a promising and see\-thing axis of current
scientific researches. Various domains of activities could benefit
from these
technologies~\cite{7044953,Brieler,Chang,QUTAC,Cheng,Sreraman}. The
domain of quantum communication has recently reached a milestone since
qubit teleportation over $1$ kilometer was
recently announced~\cite{Lago}. One of the principal goals pursued is
the construction of large quantum computers that could perform
algorithms and computations with an exponential speedup with respect
to classical computers. This perspective encounters two problems.
The
first comes from physical construction. The current effective
realizations of quantum computers host a relatively modest number of
processing elements. These elements are made of qubits which are
entities ruled by quantum mechanics rather than classical
mechanics. However, in the future, we could hope that the number of
integrated qubits grows by several magnitudes, as it has been the case
for transistors in classical computers.
The second problem is the
decoherence of qubits. Qubits have a high susceptibility to be
affected by various external perturbations. The result is that their
internal status could be rapidly affected. This feature is a major
obstacle for a faithful use of quantum computers. However, it has been
shown that if the number of altered qubits is beyond a limit, faithful
computations can be performed. Therefore, quantum error correcting
codes have been introduced in order to combat the decoherence
problem~\cite{Gott}. Since then, numerous constructions of quantum
error correcting codes have been developed. In some particular cases, efficient techniques
have also been proposed for the correction of these codes.
In this article, new quantum CSS codes and new procedures of decoding
are presented. In section~\ref{sec2}, the basics of the quantum
transmission channel are recalled. Section~\ref{sec3} is devoted to
the formalism of stabilizer and CSS codes. Section~\ref{sec4}
describes constructions involving classical codes to build new quantum
CSS codes. In section~\ref{sec5}, a new decoding procedure is
presented and applied to the classical Reed-Muller and BCH
codes. Section~\ref{sec6} presents the general framework of finite
projective geometry codes and new quantum
CSS codes are obtained. Finally section~\ref{sec7} gathers numeric
results obtained with the construction of sections~\ref{sec5} and~\ref{sec6}.
\section{Quantum transmission channel}\label{sec2}
We first introduce the basic notions and properties
of quantum objects.
\begin{definition}[Qubit]
  A qubit is a vector $(\alpha,\beta) \in \mathbb{C}^2$ such that
  $|\alpha|^2+|\beta|^2=1$.
\end{definition}
\begin{definition}[Quantum state]
  A quantum state of length $n$ is a vector
  $(\alpha_1,\alpha_2,...,\alpha_{2^n}) \in \mathbb{C}^{2^n}$ such that
  $\sum_{i=1}^{2^n}|\alpha_i|^2=1$.
\end{definition}
Therefore a qubit is a quantum state of length $1$. Some particular
quantum states need to be introduced. We will use the notations
$\ket{0}=(1,0)$ and $\ket{1}=(0,1)$. 
Therefore we have
$(\alpha,\beta)=\alpha \ket{0}
+ \beta \ket{1}$ and an arbitrary qubit is a weighted sum of $\ket{0}$ and
$\ket{1}$ with norm equal to $1$. Moreover, the quantum state of length
$n$ denoted $\ket{a_1 a_2 ... a_n}$ with $a_i\in \{0,1\}$ will be the
tensor product $\ket{a_1} \otimes \ket{a_2} \otimes ... \otimes
\ket{a_n}$. Thus,
an arbitrary quantum state of length $n$ is also a weighted sum of the
$2^n$ preceding quantum states with a norm equal to $1$.
\begin{definition}[Pauli group]
  The following four matrices of $\mathbb{C}^{2\times 2}$ are called
  Pauli matrices:
  \begin{equation}\nonumber
\begin{split}
    I = \begin{pmatrix} 1 & 0 \\0 & 1 \end{pmatrix},
    X = \begin{pmatrix} 0 & 1 \\1 & 0 \end{pmatrix},\\
    Y = \begin{pmatrix} 0 & -i \\i & 0 \end{pmatrix},
    Z = \begin{pmatrix} 1 & 0 \\0 & -1 \end{pmatrix}.
\end{split}
    \end{equation}
The Pauli group of size $n$ is then defined as:
  \begin{equation}\nonumber
    \mathcal{E}_n = \{1, -1, i, -i \} \times \{ I, X, Y, Z\}^{\otimes n}.
  \end{equation}
  Therefore, for $E\in \mathcal{E}_n$, we may write:
  \begin{equation}\nonumber
    E = \lambda \bigotimes_{i=1}^n E(i)
  \end{equation}
  with $\lambda \in  \{1, -1, i, -i \}$ and $E(i) \in  \{I, X, Y, Z
  \}$. We will also write $E= \lambda [E(1) E(2) ... E(n)]$.
\end{definition}
It can be checked that $XZ=-ZX=-iY$, $XY=-YX=iZ$, $YZ=-ZY=iX$ and
$X^2=Y^2=Z^2=I$. Therefore, two Pauli matrices $E_1$ and $E_2$ either
commute if $E_1 E_2 = E_2 E_1$ or anticommute if $E_1 E_2 = - E_2
E_1$. More precisely, a Pauli matrix commutes with itself and matrix
$I$ and anticommutes with the two other Pauli matrices.
Therefore similarly two elements of the Pauli group of size $n$ either
commute or anticommute.
\begin{definition}\label{def4}
  Let $E=[E(1) E(2) ... E(n)]$ and $F=[F(1) F(2)
    ... F(n)]$, we will note $E \bigstar F=0$ (resp. $1$) if $EF=FE$
  (resp. $EF=-FE$) and we have $E \bigstar F=\sum_{i=1}^n E(i)\bigstar
  F(i)\pmod 2$.
\end{definition}
Quantum communication channels may affect transmitted quantum states
in various ways. In order to evaluate quantum error correcting codes,
two basic channels are defined.
\begin{definition}[Pauli communication channel]
  Let $\ket{\phi}$ be a quantum state of size $n$. A Pauli communication
  channel, defined by four positive parameters $p_I$, $p_X$, $p_Y$
  and $p_Z$ such that $p_I+p_X+p_Y+p_Z=1$, outputs the quantum state
  $E\ket{\phi}$ with $E=[E(1)E(2)...E(n)]$ where for all $1\leq i \leq
  n$, $E(i)=I$ (resp, $X$, $Y$, $Z$) with probability $p_I$
  (resp. $p_X$, $p_Y$, $p_Z$).
\end{definition}
\begin{definition}[Depolarizing channel]
  A depolarizing channel, defined by a parameter $p$ with $0 \leq p
  \leq 1$, is a Pauli communication channel with $p_I=1-p$ and
  $p_X=p_Y=p_Z=p/3$.
\end{definition}
\section{Stabilizer and CSS codes}\label{sec3}
In classical communications, error correcting co\-des are used to combat
transmission errors due to the communication channel. In the framework
of quantum communications, quantum error correcting codes are also
used to deal with the effects of quantum channels. An important family
of quantum error correcting codes is the class of stabilizer
codes~\cite{Gott}.
\begin{definition}[Stabilizer code]
  Let $\mathcal{S}$ be a commutative subgroup of $\mathcal{E}_n$ not
  containing $-I^{\otimes n}$, then the quantum error correcting code
  $\mathcal{C}$ stabilized by $\mathcal{S}$ is defined as:
    \begin{align*}
    \mathcal{C} = & \{\ket{\phi}\in \mathbb{C}^{2^n} \text{such that } S\ket{\phi}=\ket{\phi} \\
    & \text{ for all } S \in\mathcal{S} \}.
    \end{align*}
\end{definition}
Classical notions for error correcting codes
can then be defined in the quantum communication framework.
\begin{definition}[Dimension of a stabilizer code]
  Let $\mathcal{C}$ be a stabilizer code of $n$ qubits. The dimension
  of $\mathcal{C}$, denoted $\mathrm{dim}(\mathcal{C})$ is the integer
  value $k$ such that $\mathcal{C}$ is a subspace of dimension $2^k$
  in $\mathbb{C}^{2^n}$.
\end{definition}
It can then be proven that:
\begin{theorem}
  Let $\mathcal{C}$ be a stabilizer code of $n$ qubits stabilized by
  $\mathcal{S}$. Let $r$ be the number of independent generators of
  $\mathcal{S}$. We have then:
  \begin{equation}\nonumber
    \mathrm{dim}(\mathcal{C}) = n-r.
  \end{equation}
\end{theorem}
\begin{definition}[Centralizer]
Let $\mathcal{S}$ be the stabilizer of a quantum code $\mathcal{C}$ of
length $n$. The centralizer of $\mathcal{S}$, denoted $C(\mathcal{S})$
is the subgroup generated by all the elements of $\mathcal{E}_n$
commuting with all the elements of $\mathcal{S}$:
\begin{align*}
    C(\mathcal{S}) = & \{ E \in \mathcal{E}_n \text{ such that } E
    \bigstar S = 0 \\
    & \text{ for all } S \in \mathcal{S} \}.
\end{align*}
\end{definition}
\begin{definition}[Syndrome]
Let $\mathcal{S}=\{ S_1,S_2,...,$ $S_r\}$ be the stabilizer of a quantum
code $\mathcal{C}$. Let $E$ be a Pauli error. The syndrome of $E$ is
then the vector $s(E)=(E\bigstar S_1,E\bigstar S_2,...,E\bigstar S_r)$.
\end{definition}
As in the classical case, syndrome values are used to detect and
eventually correct transmission errors. Suppose that $E\bigstar S_i=1$
for some $1\leq i \leq r$. We have then, for any $\ket{\phi}\in
\mathcal{C}$, $E(\ket{\phi})=ES_i(\ket{\phi})=-S_iE(\ket{\phi})$, therefore
$E(\ket{\phi})$ is not stabilized by $S_i$ and thus is not a codeword of
$\mathcal{C}$. Therefore an error is detected. Suppose now that $E\in
\mathcal{S}$. Then by definition, $E(\ket{\phi})=\ket{\phi}$ for all
$\ket{\phi}\in \mathcal{C}$. We have then a benign error since it does not
affect the codewords of $\mathcal{C}$. In this case, we have also
$s(E)=0$ and, in fact, this error is not detected.
In the third case, we
have $E\in C(\mathcal{S}) \setminus \mathcal{S}$. We have then
$s(E)=0$ and
this error is not detected.
However, by definition of
$\mathcal{C}$ and since $E\not\in
\mathcal{S}$, there is a codeword $\ket{\phi}\in \mathcal{C}$ such that
$E(\ket{\phi}) \neq \ket{\phi}$. The code $\mathcal{C}$ is then not invariant on
the action of $E$. As a consequence, the received quantum state $E(\ket{\phi)}$ is 
potentially erroneous and undetected. It is said that a serious error
has occurred. A consequence of this particularity is that for a
quantum code $\mathcal{C}$, two kinds of minimum distance are defined.
\begin{definition}[Minimum distance]
  Let $\mathcal{C}$ be a quantum code stabilized by $\mathcal{S}$. The
  minimum distance of $\mathcal{C}$ is then defined by:
  \begin{equation}\nonumber
    d_{min}(\mathcal{C}) = \mathrm{min}\{ w(E) \text{ such that } E\in
    C(S)\setminus S\}
  \end{equation}
  where $w(E)=\sum_{i=1}^n \mathbbm{1}(E(i)\neq I)$.
\end{definition}
\begin{definition}{\bf(Non-degenerate minimum distance)}
  Let $\mathcal{C}$ be a quantum code stabilized by $\mathcal{S}$. The
  non-degenerate minimum distance of $\mathcal{C}$ is then defined by:
  \begin{equation}\nonumber
    \begin{split}
    d'_{min}(\mathcal{C}) = \mathrm{min}\{ &w(E) \text{ such that } E\in
    C(S)\\
    & \text{ and } E\neq I^{\otimes n}\}.
    \end{split}
  \end{equation}
\end{definition}
We can now introduce the Calderbank-Shor-Steane
construction of quantum error correcting codes~\cite{CS,Steane}.
\begin{definition}[CSS quantum code]\label{CSS}
  Let $\mathsf{C_1}$ (resp. $\mathsf{C_2}$) be a classical binary
  linear code of length $n$ and 
  dimension $k_1$ (resp. $k_2$) such that
  $\mathsf{C_2}\subset \mathsf{C_1}^\perp$. We set
  $d_1=d_{min}(\mathsf{C_1^\perp})$, $d_2=d_{min}(\mathsf{C_2^\perp})$
  and $d=\mathrm{min}(d_1,d_2)$.
  Let $G_1$ (resp. $G_2$) be a $k_1\times n$ (resp. $k_2\times n$)
  generating 
  matrix of $\mathsf{C_1}$ (resp. $\mathsf{C_2}$).
  Let $\mathcal{S}_X=\{ S_{X,1}, S_{X,2}, ... ,S_{X,k_1} \}$
  (resp. $\mathcal{S}_Z$) such that
  $S_{X,i} = \bigotimes_{j=1}^n X^{G_{1ij}}$ (resp. $S_{Z,i} = \bigotimes_{j=1}^n
  Z^{G_{2ij}}$). Then the code $\mathcal{C}$ stabilized by
  $\mathcal{S}=\mathcal{S}_X \cup \mathcal{S}_Z$ is a $[[n,n-(k_1+k_2),d]]$
  quantum code i.e. of
  length $n$, dimension $n-(k_1+k_2)$ and minimum distance $d$.

  If $\mathfrak{c}$ is a codeword of $\mathcal{C}$, then it is of the
  form:
  \begin{equation}\nonumber
  \begin{split}
    \mathfrak{c} & =\frac{1}{2^{k_2/2}} \sum_{w\in C_2} \ket{c+w}
    \end{split}
  \end{equation}
  for $c$ being a codeword of $\mathsf{C_1}^\perp$.
\end{definition}

A classical particular case of CSS codes is given by the following
additional hypothesis: $\mathsf{C_1}=\mathsf{C_2}$.
A code $\mathsf{C}$ such that $\mathsf{C}\subset
\mathsf{C}^\perp$ is said to be self-orthogonal or weakly
self-dual. For instance, if $\mathsf{C}$ is a subcode of a
self-dual code, then it is a self-orthogonal code. For example, the
$7$-qubit Steane code can be defined in the stabilizer formalism as
the CSS quantum code defined from the self-orthogonal $[7,3,4]$ code
whose dual is
the Hamming
$[7,4,3]$ code.
\section{New self-orthogonal codes from old ones}\label{sec4}
In~\cite[Ch. 1 {\S}9, Ch. 2 {\S}9, Ch. 18]{Sloane}, many constructions
are given to construct new codes from old ones. In this section, we
review these methods and several from other sources from the point of
view of self-orthogonality,
in the prospect of
building new quantum codes.
\begin{theorem}[Code augmentation]
  Let $\mathsf{C}$ be a binary self-orthogonal linear code of length $n$ and dimension
  $k$ such that $\mathrm{d}_\mathrm{min} (\mathsf{C}^\perp) = d$. If $n$
  is even and $\mathsf{C}$ does not contain the length $n$ all one
  codeword $\mathbf{1}$, then the code $\mathsf{C'}=\mathsf{C} \cup
  \{\mathbf{1}+ \mathsf{C}\}$ is self-orthogonal and we
  have $d'=\mathrm{d}_\mathrm{min} (\mathsf{C'}^\perp) \geq d$. The
  corresponding CSS quantum code is a $[[n,n-2k-2,d']]$
  quantum code.
\end{theorem}
\paragraph{Proof :}
  Let $c'_1$ and $c'_2$ two codewords of $\mathsf{C'}$. We have then
  $c'_1=\lambda \mathbf{1} + c_1$ and $c'_2=\mu \mathbf{1} + c_2$ with
  $c_1,c_2 \in \mathsf{C}$ and $\lambda, \mu \in \{0,1\}$. Thus the
  scalar product of $c_1$ and $c_2$ is :
  \begin{align}
    c'_1.c'_2 & =(\lambda \mathbf{1} + c_1).(\mu \mathbf{1} +
    c_2)\nonumber \\
    & = \lambda \mu n \pmod 2\nonumber \\
    & + \lambda w_H(c_2) \pmod 2\nonumber \\
    & + \mu w_H(c_1) \pmod 2\nonumber \\
    & + c_1.c_2.\nonumber 
  \end{align}
  where $w_H$ denotes the Hamming weight function.
  The first right-hand side term is null since $n$ is even and the
  three other right-hand side terms are null since $\mathsf{C}$ is self
  orthogonal. Therefore $\mathsf{C'}$ is also a self orthogonal code.
  Moreover $\mathsf{C} \subset \mathsf{C'}$, so that
  $\mathsf{C'}^\perp \subset \mathsf{C}^\perp$ and thus
  $\mathrm{d}_\mathrm{min} (\mathsf{C'}^\perp) \geq \mathrm{d}_\mathrm{min}
  (\mathsf{C}^\perp)$.
$\phantom{gauche} \qedwhite$
  \begin{theorem}[Code shortening]
    Let $\mathsf{C}$ be a binary self-orthogonal linear code of length
    $n$ and dimension
  $k$ such that $\mathrm{d}_\mathrm{min} (\mathsf{C}^\perp) = d$. Let
    $1\leq i\leq n$ and:
    \begin{align}
      \mathsf{C'} & =\{ (c'_1,c'_2,...,c'_{i-1},c'_{i+1},...,c'_n)\nonumber \\
      & \text{ } \mathrm{such} \text{ } \mathrm{that} \text{ }
      (c'_1,c'_2,...,c'_{i-1},0,c'_{i+1},...,c'_n)\in
      \mathsf{C} \}.\nonumber 
    \end{align}
    Then $\mathsf{C'}$ is a binary self-orthogonal linear code of length
    $n-1$ and dimension
    $k-1$ and we have $\mathrm{d}_\mathrm{min} (\mathsf{C'}^\perp) \geq d-1$.
  \end{theorem}
  \paragraph{Proof :}
  The fact that $\mathsf{C'}$ is self-orthogonal is obvious.
  The code $\mathsf{C'}$ has been constructed from the codewords of
  $\mathsf{C}$ having $0$ at their $i$-th coordinate.
  Therefore $\mathsf{C'}$ contains half the number of codewords of
  $\mathsf{C}$ and its dimension is equal to $k-1$.
  Thus the
  dimension of $\mathsf{C'}^\perp$ is equal to $n-k$ and is equal to
  the dimension of $\mathsf{C}^\perp$. Moreover every codeword of
  $\mathsf{C}^\perp$ punctured at the $i$-th bit belongs to
  $\mathsf{C'}^\perp$. Therefore $\mathsf{C'}^\perp$ contains only
  this type of codewords and $\mathrm{d}_\mathrm{min} (\mathsf{C'}^\perp)
  \geq d-1$.
  $\phantom{gauche} \qedwhite$
  \begin{example}[Steane 7-qubit code]~\label{stea9}
    Let $\mathsf{C}$ be the extended Hamming $[8,4,4]$ self-dual code. If
    we shorten it, we obtain a $[7,3,4]$ self-orthogonal code, whose
    dual has parameters $[7,4,3]$. With the CSS construction, we
    obtain then a $[[7,1,3]]$ quantum code, which is in fact the
    Steane 7-qubit code. This quantum error correcting code is optimal in
    minimum distance amongst the CSS codes according to the tables of
    best classical error correcting codes~\cite{681315,Grassl:codetables}.
    However, there is also a shorter
    $[[5,1,3]]$ stabilizer code. Likewise, if $n=4k+2$
    with $n-1$ being a prime number, the extended quadratic residue
    code of length $n$ is self-dual and its minimum distance is
    greater than $\sqrt{n}$. By shortening, it is then possible to
    define an infinite family of quantum codes with dimension $1$,
    whose minimum distances grow like the square root of their length.
  \end{example}
  \begin{theorem}[Plotkin construction]\label{plotk}
    Let $\mathsf{C_1}$ and $\mathsf{C_2}$ be two binary self-orthogonal linear codes of length
    $n$ and respective dimension
    $k_1$ and $k_2$ such that $\mathsf{C_2}\subset \mathsf{C_1}^\perp$.
    Then the code $\mathsf{C'}$ defined as:
      \begin{align}
      \mathsf{C'} =\{ (u|u+v) \text{ } \mathrm{s.t.} \text{ } u\in \mathsf{C_1}
      \text{ } \mathrm{and} \text{ } v\in \mathsf{C_2} \} \nonumber
      \end{align}
      is self-orthogonal code of length $2n$ and dimension $k_1+k_2$.
      We also have:
      \begin{align}
      \mathsf{C'^\perp} =\{ (u^*+v^*|v^*) \text{ } \mathrm{s.t.} \text{ } u^*\in \mathsf{C_1^\perp}
      \text{ } \mathrm{and} \text{ } v^*\in \mathsf{C_2^\perp} \}. \nonumber
      \end{align}
      Moreover, if $d_1=\mathrm{d}_\mathrm{min} (\mathsf{C_1}^\perp)$ and
      $d_2=\mathrm{d}_\mathrm{min} (\mathsf{C_2}^\perp)$ then
      $\mathrm{d}_\mathrm{min} (\mathsf{C'}^\perp)=\mathrm{min}(2d_2,d_1)$.
  \end{theorem}
  \paragraph{Proof :}
  Let $(u_1|u_1+v_1)$ and $(u_2|u_2+v_2)$ be two codewords of
  $\mathsf{C'}$. Then we have:
  \begin{align}
    (u_1|u_1+v_1).&(u_2|u_2+v_2) \nonumber \\
    &= u_1.u_2 + (u_1+v_1).(u_2+v_2)\nonumber \\
    &= u_1.v_2+u_2.v_1=0.\nonumber
  \end{align}
  since $\mathsf{C_1}$ and $\mathsf{C_2}$ are self-orthogonal and
  $\mathsf{C_2}\subset \mathsf{C_1}^\perp$.
  Let $(u|u+v)$ be a codeword of $\mathsf{C'}^\perp$ and let
  $(u^*+v^*|v^*)$ be a codeword with $u^*\in \mathsf{C_1}^\perp$ and
  $v^*\in \mathsf{C_2}^\perp$. We have then:
  \begin{align}
    (u|u+v).&(u^*+v^*|v^*)\nonumber \\
    &= u.u^*+u.v^*+u.v^*+v.v^*=0.\nonumber 
  \end{align}
  The dimension of $\mathsf(C')^\perp$ is equal to $2n-(k_1+k_2)$.
  The dimension of the code generated by the codewords of the form
  $(u^*+v^*|v^*)$ is equal to
  $\mathrm{dim}(\mathsf(C_1)^\perp)+\mathrm{dim}(\mathsf(C_2)^\perp)=
  (n-k_1)+(n-k_2)$. Both values are equal and therefore we have:
  \begin{align}
    \mathsf{C'}^\perp=\{ (u^*+v^*|v^*) \text{ } \mathrm{s.t.} \text{ } u\in \mathsf{C_1}^\perp
      \text{ } \mathrm{and} \text{ } v\in \mathsf{C_2}^\perp \} \nonumber.
  \end{align}
  The code $\mathsf{C'^\perp}$ is then obtained from a Plotkin
  construction involving $\mathsf{C_1^\perp}$ and $\mathsf{C_2^\perp}$
  and a permutation of the two contiguous length $n$ blocks of the
  codewords. Then by~\cite[th. 33 p. 76]{Sloane}, we have
  $\mathrm{d}_\mathrm{min} (\mathsf{C'}^\perp)=\mathrm{min}(2d_2,d_1)$.
  $\phantom{gauche} \qedwhite$
  \begin{example}
    The family of Reed-Muller codes that will be addressed in
    paragraph~\ref{qrm}
    can be built recursively by the Plotkin construction. 
  \end{example}
  
  \begin{theorem}
    Let $\mathsf{C_1}$ and $\mathsf{C_2}$ be two binary self-orthogonal linear codes of length
    $n$ and respective dimension
    $k_1$ and $k_2$ such that $\mathsf{C_2}\subset \mathsf{C_1}^\perp$.
    Then the code $\mathsf{C'}$ defined as:
      \begin{align}
      \mathsf{C'} =\{ (u+w|v+w|&u+v+w) \text{ } \mathrm{s.t.} \text{ } u,v\in
      \mathsf{C_1} \nonumber \\
      &\text{ } \mathrm{and} \text{ } w\in \mathsf{C_2} \} \nonumber
      \end{align}
      is self-orthogonal code of length $3n$ and dimension $2.k_1+k_2$.
      We also have:
      \begin{align}
        \mathsf{C'^\perp} & =\{ (u^*+w^*|v^*+w^*|u^*+v^*+w^*) \nonumber \\
        &\text{ } \mathrm{s.t.}\text{ } 
        u^*,v^*\in \mathsf{C_1}^\perp
        \text{ } \mathrm{and} \text{ } w^*\in \mathsf{C_2}^\perp \}. \nonumber
      \end{align}
  \end{theorem}
  \paragraph{Proof :}
  The proof is similar to that of Theorem~\ref{plotk} and is therefore
  skipped.  However, in this case, we cannot obtain an estimate of the minimum
  distance of $\mathsf{C'^\perp}$.
  $\phantom{gauche} \qedwhite$
  
  The next construction was proposed
  in~\cite{Nebe} for self-dual codes and generalizes the Turyn
  construction of the extended Golay code.
  \begin{theorem}[Nebe construction]
    Let $\mathsf{C}$ and $\mathsf{D}$ be two binary self-orthogonal
    linear codes
    of length
    $n$ and dimension
    $k$ and $\mathsf{E}$ a binary linear code of length $m$.
    Then the
    code $\mathsf{C}\otimes \mathsf{E}+\mathsf{D}\otimes
    \mathsf{E}^\perp$ is a binary self-orthogonal
    linear code of length $nm$ and dimension $km$.
  \end{theorem}
  \paragraph{Proof :}
  Let $c,c'\in \mathsf{C}$, $d,d'\in \mathsf{D}$, $e,e'\in
  \mathsf{E}$ and $f,f'\in
  \mathsf{E}^\perp$. We have then:
  \begin{align}
    (c\otimes e).(c'\otimes e')=0 & \text{ }\mathrm{since} \text{ }
    \mathsf{C}\subset \mathsf{C}^\perp\mathrm{,}\nonumber\\
    (d\otimes f).(d'\otimes f')=0 & \text{ }\mathrm{since} \text{ }
    \mathsf{D}\subset \mathsf{D}^\perp\mathrm{,\nonumber}\\
    (c\otimes e).(d\otimes f)=0 & \text{ }\mathrm{since} \text{ }
    e\in \mathsf{E}\text{ }\mathrm{and} \text{ }f\in \mathsf{E}^\perp\mathrm{.}\nonumber
  \end{align}
  $\phantom{gauche} \qedwhite$
  \begin{theorem}[Product codes]
    Let $\mathsf{C_1}$ and $\mathsf{C_2}$ be two binary linear codes
    of respective lengths
    $n_1$ and $n_2$, and respective dimension
    $k_1$ and $k_2$. Let $\mathsf{C'}=C_1\otimes C_2$ be the product
    code of $\mathsf{C_1}$ and $\mathsf{C_2}$. The code $\mathsf{C'}$
    has a length of $n_1n_2$ and a dimension of $k_1k_2$. Then if at
    least one code amongst $C_1$ and $C_2$ is self-orthogonal, then
    $\mathsf{C'}$ is also self-orthogonal.
    
    Moreover, if $d_1=\mathrm{d}_\mathrm{min} (\mathsf{C_1}^\perp)$ and
    $d_2=\mathrm{d}_\mathrm{min} (\mathsf{C_2}^\perp)$ then
    $\mathrm{d}_\mathrm{min} (\mathsf{C'}^\perp)=\mathrm{min}(d_1,d_2)$.
  \end{theorem}
  \paragraph{Proof :}
  Let $c'\in\mathsf{C'}$. As usually $c'$ will be denoted as a
  $n_1\times n_2$ binary matrix such that for all $1\leq i \leq n_1$,
  the vector $(c'_{i1},c'_{i2},...,c'_{in_2})$ is a codeword of
  $\mathsf{C_2}$ for all $1\leq j \leq n_2$,
  the vector $(c'_{1j},c'_{2j},...,c'_{n_1 j})$ is a codeword of
  $\mathsf{C_1}$.
  Let $c^*$ be a binary word of length $n_1n_2$ represented by an
  $n_1\times n_2$ matrix. The matrix is built such that any of its
  columns contains an arbitrary codeword of $\mathsf{C_1}^\perp$. We
  have then:
  \begin{align}\nonumber
    c'.c^* = \sum_{j=1}^{n_2} \sum_{i=1}^{n_1} c'_{ij}c^*_{ij}=0 
  \end{align}
  since columns by columns we have the dot product of a codeword of
  $\mathsf{C_1}$ and $\mathsf{C_1}^\perp$. We can then build $2^{(n_1-k_1)*n_2}$
  distinct codewords of $\mathsf{C'}^\perp$.
  Likewise, if we build $c^*$ row by row using codewords of
  $\mathsf{C_2}^\perp$, we obtain another set of $2^{(n_2-k_2)*n_1}$
  codewords of $\mathsf{C'}^\perp$ distinct with each other.
  However both sets are not disjoints. They both contain the
  codewords $c^*$ where any row is a codeword of $\mathsf{C_2}^\perp$
  and any column is a codeword of $\mathsf{C_1}^\perp$. This case is
  the case of codewords $c^*$ belonging to $\mathsf{C_1}^\perp \otimes
  \mathsf{C_2}^\perp$.
  All these sets are linear, therefore the set of all codewords of
  $\mathsf{C'}^\perp$ generated this way has a dimension equal to
  $(n_1-k_1)n_2+(n_2-k_2)n_1-(n_1-k_1)(n_2-k_2)=n_1n_2-k_1k_2$. Since
  the dimension of $\mathsf{C'}$ is equal to $k_1 k_2$ and its length
  is $n_1n_2$, we have obtained an exact description
  of $\mathsf{C'}^\perp$. Finally, if we consider codewords $c^*$
  with a unique non null row or a unique non null column, we
  conclude that $\mathrm{d}_\mathrm{min}(\mathsf{C'}^\perp)=\mathrm{min}(d_1,d_2)$.
$\phantom{gauche} \qedwhite$

\begin{figure}[htbp]
  \begin{center}
\begin{picture}(0,0)%
\includegraphics{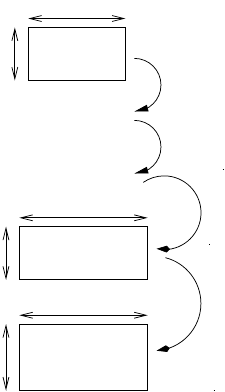}%
\end{picture}%
\setlength{\unitlength}{3108sp}%
\begingroup\makeatletter\ifx\SetFigFont\undefined%
\gdef\SetFigFont#1#2#3#4#5{%
  \reset@font\fontsize{#1}{#2pt}%
  \fontfamily{#3}\fontseries{#4}\fontshape{#5}%
  \selectfont}%
\fi\endgroup%
\begin{picture}(2277,3972)(2146,-3808)
\put(3871,-736){\makebox(0,0)[lb]{\smash{{\SetFigFont{9}{10.8}{\rmdefault}{\mddefault}{\updefault}{\color[rgb]{0,0,0}$\varphi$}%
}}}}
\put(2971,-1681){\makebox(0,0)[b]{\smash{{\SetFigFont{12}{14.4}{\rmdefault}{\mddefault}{\updefault}{\color[rgb]{0,0,0}$(B_1,...,B_{n_2})$}%
}}}}
\put(3871,-1366){\makebox(0,0)[lb]{\smash{{\SetFigFont{9}{10.8}{\rmdefault}{\mddefault}{\updefault}{\color[rgb]{0,0,0}$\mathsf{C_2}$}%
}}}}
\put(2926,-1051){\makebox(0,0)[b]{\smash{{\SetFigFont{12}{14.4}{\rmdefault}{\mddefault}{\updefault}{\color[rgb]{0,0,0}$(B_1,...,B_{k_2})$}%
}}}}
\put(2251,-466){\makebox(0,0)[rb]{\smash{{\SetFigFont{9}{10.8}{\rmdefault}{\mddefault}{\updefault}{\color[rgb]{0,0,0}$k_1$}%
}}}}
\put(2926,-466){\makebox(0,0)[b]{\smash{{\SetFigFont{9}{10.8}{\rmdefault}{\mddefault}{\updefault}{\color[rgb]{0,0,0}$A_{i,j}$}%
}}}}
\put(2791, 29){\makebox(0,0)[lb]{\smash{{\SetFigFont{9}{10.8}{\rmdefault}{\mddefault}{\updefault}{\color[rgb]{0,0,0}$k_2$}%
}}}}
\put(2161,-2491){\makebox(0,0)[rb]{\smash{{\SetFigFont{9}{10.8}{\rmdefault}{\mddefault}{\updefault}{\color[rgb]{0,0,0}$k_1$}%
}}}}
\put(3016,-2446){\makebox(0,0)[b]{\smash{{\SetFigFont{9}{10.8}{\rmdefault}{\mddefault}{\updefault}{\color[rgb]{0,0,0}$A_{i,j}$}%
}}}}
\put(2836,-1996){\makebox(0,0)[lb]{\smash{{\SetFigFont{9}{10.8}{\rmdefault}{\mddefault}{\updefault}{\color[rgb]{0,0,0}$n_2$}%
}}}}
\put(4276,-2041){\makebox(0,0)[lb]{\smash{{\SetFigFont{9}{10.8}{\rmdefault}{\mddefault}{\updefault}{\color[rgb]{0,0,0}$\varphi^{-1}$}%
}}}}
\put(4276,-2986){\makebox(0,0)[lb]{\smash{{\SetFigFont{9}{10.8}{\rmdefault}{\mddefault}{\updefault}{\color[rgb]{0,0,0}$\mathsf{C_1}$}%
}}}}
\put(2206,-3571){\makebox(0,0)[rb]{\smash{{\SetFigFont{9}{10.8}{\rmdefault}{\mddefault}{\updefault}{\color[rgb]{0,0,0}$n_1$}%
}}}}
\put(3061,-3526){\makebox(0,0)[b]{\smash{{\SetFigFont{9}{10.8}{\rmdefault}{\mddefault}{\updefault}{\color[rgb]{0,0,0}$A_{i,j}$}%
}}}}
\put(2836,-2986){\makebox(0,0)[lb]{\smash{{\SetFigFont{9}{10.8}{\rmdefault}{\mddefault}{\updefault}{\color[rgb]{0,0,0}$n_2$}%
}}}}
\end{picture}%
    \caption{Encoding procedure of $\mathsf{C_1}\bigstar_{\varphi}
    \mathsf{C_2}$\label{encconc}}
  \end{center}
\end{figure}

  In what concerns concatenated codes, we will refer to the definition
  given in~\cite[Ch. 10 p. 307]{Sloane}.
  \begin{definition}[Concatenated codes]
  Let $\mathsf{C_1}$ be a binary linear code
    of length
    $n_1$ and dimension $k_1$, $\mathsf{C_2}$ a linear code of length
    $n_2$ and dimension $k_2$ with
    coefficients in $\mathrm{GF}(2^{k_1})$ and $\varphi$ a linear
    bijective mapping
    from $\mathrm{GF}(2)^{k_1}$ to $\mathrm{GF}(2^{k_1})$.
    We suppose, without loss of generality, that $\mathsf{C_1}$ and
    $\mathsf{C_2}$ are given in systematic form i.e.,
    the first $k_1$ bits (resp. $k_2$ symbols in
    $\mathrm{GF}(2^{k_2})$) of
    codewords of $\mathsf{C_1}$ (resp. $\mathsf{C_2}$) is an information part.
    Then the
    concatenated code $\mathsf{C'}=\mathsf{C_1}\bigstar_{\varphi} \mathsf{C_2}$
    is defined as the set of codewords of the form
    $(A_1,A_2,...,A_{n_2})$ where the $A_j$ are codewords of
    $\mathsf{c_1}$ and such that $(B_1,B_2,...,B_{k_2})$ is a codeword
    of $\mathsf{C_2}$, if you set
    $B_j=\varphi(A_{1,j},A_{2,j},...,A_{k_1},j)$ for all $1\leq i \leq
    k_2$. The
    dimension of $\mathsf{C_1}\bigstar_{\varphi} \mathsf{C_2}$ is then
    $k_1 k_2$ and the length of the codewords is $n_1 n_2$.
  \end{definition}
  Therefore the encoding of $\mathsf{C_1}\bigstar_{\varphi}
  \mathsf{C_2}$ is made of two steps. The set of bits to be encoded is
  the set $A_{i,j}$ with $1\leq i \leq k_1$ and $1\leq j \leq k_2$. The
  first encoding begins by the projection of the $k_1$-tuples
  $(A_{1,j},A_{2,j},...,A_{k_1,j})$ with $1\leq j \leq k_2$ by $\varphi$
  and gives the $k_2$ symbols $(B_1,B_2,...,B_{k_2})$ which are then
  encoded by the encoding matrix of $\mathsf{C_2}$ to obtain the $n_2$
  symbols $(B_1,B_2,...,B_{n_2})$. Using the inverse projection
  $\varphi^{-1}$, the set of bits $A_{i,j}$ with $1\leq i \leq k_1$
  and $1\leq j \leq n_2$ is then obtained. Then the second step of
  encoding consists in encoding this new set columnwise by the
  encoding function of $\mathsf{C_1}$ to obtain the entire frame
  $A_{i,j}$ with $1\leq i \leq n_1$ and $1\leq j \leq n_2$. The
  encoding procedure is illustrated in figure~\ref{encconc}.
  We have then:
  \begin{theorem}[Concatenated codes]
    Let $\mathsf{C_1}$ be a binary linear code
    of length
    $n_1$ and dimension $k_1$, $\mathsf{C_2}$ a linear code of length
    $n_2$ and dimension $k_2$ with
    coefficients in $\mathrm{GF}(2^{k_1})$ and $\varphi$ a linear bijective mapping
    from $\mathrm{GF}(2)^{k_1}k$ to $\mathrm{GF}(2^{k_1})$. If $\mathsf{C_1}$
    is a self-orthogonal code, then
    $\mathsf{C'}=\mathsf{C_1}\bigstar_{\varphi} \mathsf{C_2}$ is also
    a self-orthogonal code. Moreover we have
    $\mathrm{d}_\mathrm{min}(\mathsf{C'}^\perp)=\mathrm{d}_\mathrm{min}(\mathsf{C_1}^\perp)$. 
  \end{theorem}
  \paragraph{Proof :}
  In figure~\ref{encconc}, all columns are codewords of $\mathsf{C_1}$.
  Thus the fact that $\mathsf{C'}$ is self-orthogonal is quite obvious.
  In the encoding process, for a given column $j$, all the codewords
  of $\mathsf{C_1}$ appear in the $j$-th column when all codewords of
  $\mathsf{C'}$ are considered.  Therefore,
  we have $\mathrm{d}_\mathrm{min}(\mathsf{C'}^\perp)\geq
  \mathrm{d}_\mathrm{min}(\mathsf{C_1}^\perp)$.
  Reciprocally let $c^*$
  be a binary $n_1 \times n_2$ matrix such that
  is such that $c^*_{\_j}\in \mathsf{C_1}$ and $c^*_{\_k}=0$ for all
  $k\neq j$, then $c^*$ has a null dot product with any codeword of
  $\mathsf{C'}$, so that $c^*\in \mathsf{C'}^\perp$. Thus, we have,
  $\mathrm{d}_\mathrm{min}(\mathsf{C'}^\perp)=\mathrm{d}_\mathrm{min}(\mathsf{C_1}^\perp)$.
  $\phantom{gauche} \qedwhite$
  \begin{theorem}[Construction X]\label{thx}
    Let $\mathsf{C_1}$ be a binary linear code
    of length
    $n_1$ and dimension $k_1$, $\mathsf{C_2}$ a binary self-orthogonal
    linear code of length 
    $n_1$ and dimension $k_2$ such that $\mathsf{C_1}\subset \mathsf{C_2}$
    and $\mathsf{C_3}$ a binary self-orthogonal
    linear code of length
    $n_3$ and dimension $k_2-k_1$.
    Let $\pi$ a linear surjective projection from
    $\mathsf{C_2}$ to $\mathsf{C_2}/\mathsf{C_1}$,
    with $\mathsf{C_1}$ and $\mathsf{C_2}$ considered as vectorial
    spaces. 
    Let
    $\varphi$ a linear bijection from $\mathsf{C_2}/\mathsf{C_1}$ to
    $\mathsf{C_3}$.
    Then the code $\mathsf{C'}=\{ (c_2|\varphi(\pi(c_2))) \text{ }
    \mathrm{s.t.} \text{ }
    c_2\in\mathsf{C_2} \}$  that we denote
    $X(\mathsf{C_1},\mathsf{C_2},\mathsf{C_3})$ is
    self-orthogonal. Moreover, if
    $d_2=\mathrm{d}_\mathrm{min}(\mathsf{C_2}^\perp)$ and
    $d_3=\mathrm{d}_\mathrm{min}(\mathsf{C_3}^\perp)$, then
    $\mathrm{d}_\mathrm{min}(\mathsf{C'}^\perp)=\mathrm{min}(d_2,d_3)$.
  \end{theorem}
  \paragraph{Proof :}
  Let $c'_1=(c_{21}|\varphi(\pi(c_{21})))$ and
  $c'_2=(c_{22}|\varphi(\pi(c_{22})))$ be two codewords of
  $\mathsf{C'}$ with $c_{21},c_{22}\in \mathsf{C_2}$. We have then:
  \begin{align}
    c'_1.c'_2 & = (c_{21}.c_{22} +
    \varphi(\pi(c_{21})).\varphi(\pi(c_{22}))) \pmod 2\nonumber\\
    & = 0 \nonumber
  \end{align}
  since $\mathsf{C_2}$ is self-orthogonal, $\mathrm{Im} (\varphi\circ
  \pi)=\mathsf{C_3}$ and $\mathsf{C_3}$ is self-orthogonal.
  Let $c^*$ be a non null codeword of $\mathsf{C'}^\perp$. The length
  of $c^*$ is equal to $n_1+n_3$. Let denote $c^*=(c^*_l|c^*_r)$ where
  $c^*_l$ (resp. $c^*_r$) has length $n_1$ (resp. $n_3$). Since
  $c^*\neq 0$, we have either $c^*_l\neq 0$ or $c^*_r\neq 0$.
  In the first case, we have necessarily $c^*_l\in \mathsf{C_2}^\perp$
  because $(c^*_l|0)\in \mathsf{C'}^\perp$.
  
  Reciprocally any codeword of the form $(c^*_l|0)$ with $c^*_l \in \mathsf{C_2}$
  belongs to
  $\mathsf{C'}^\perp$. In the second case, we have $c^*_r\in
  \mathsf{C_3}^\perp$ and $(0|c^*_r)\in \mathsf{C'}^\perp$. Therefore,
  we have $\mathrm{d}_\mathrm{min}(\mathsf{C'}^\perp)=\mathrm{min}(d_2,d_3)$.
  $\phantom{gauche} \qedwhite$
  
  In the preceding encoding procedure $\mathsf{C_2}$ is partitioned as cosets
  of $\mathsf{C_1}$ with $\mathsf{C_2}=\cup_{1\leq k\leq k_2-k_1}
  \{d_k+\mathsf{C_1} \}$, with $d_1$ is the all null codeword of
  length $n_1$. The function $\pi$ is then the canonical
  projection, say that if $c\in \{ d_k+\mathsf{C_1}\}$, then
  $\pi(c)=d_k$. This function is chosen to be linear, so that if
  $\pi(c_1+c_2)= \pi(c_1)+\pi(c_2)$. The set of cosets leader is then
  put in bijection with codewords of $\mathsf{C_3}$ by the function
  $\phi$. The codewords are finally formed as the concatenation of a
  codeword $c\in \mathsf{C_2}$ and its image in $\mathsf{C_3}$ by
  $\phi \circ \pi$.
  \begin{example}
    As we will see in paragraph~\ref{bch}, the dual codes of the
    $[31,26,3]$ and the $[31,21,5]$ BCH
    codes are self-orthogonal. 
    These dual codes $\mathsf{C_1}$ and $\mathsf{C_2}$ have parameters
    $[31,5]$ and $[31,10]$. 
    Moreover the $[31,21,5]$ BCH code is included in the $[31,26,3]$
    BCH code and thus the $[31,5]$ dual BCH code is included in the
    $[31,10]$ dual BCH code. Moreover the $\mathsf{C_3}=[16,5,8]$
    Reed-Muller code
    is self orthogonal and its dual is the $[16,11,4]$ Reed-Muller
    code. By construction of
    $X(\mathsf{C_1},\mathsf{C_2},\mathsf{C_3})$,
    we obtain a self-orthogonal of length $47$ and dimension $10$,
    whose dual code has a minimum distance
    of at least $4$. Therefore we can build a $[[47,27,4]]$ CSS
    quantum code.
  \end{example}
  \begin{theorem}[Construction X3]
    Let $\mathsf{C_1}$,  $\mathsf{C_2}$ and $\mathsf{C_3}$ be three binary
    self-orthogonal linear codes
    of length
    $n_1$, respective dimensions $k_1$, $k_2$ and $k_3$ such that
    $\mathsf{C_1}\subset \mathsf{C_2}\subset \mathsf{C_3}$. Let
    $\pi_2$ and $\pi_3$ be projections respectively, from
    $\mathsf{C_2}$ to $\mathsf{C_2}/\mathsf{C_1}$ and from
    $\mathsf{C_3}$ to $\mathsf{C_3}/\mathsf{C_2}$. Let $\mathsf{C_4}$
    and $\mathsf{C_5}$ be two binary
    self-orthogonal linear codes
    of respective lengths $n_4$ and $n_5$ and respective
    dimensions $k_2-k_1$ and $k_3-k_2$. Let $\varphi_2$ and
    $\varphi_3$ two linear bijections respectively from
    $\mathsf{C_2}/\mathsf{C_1}$ to $\mathsf{C_4}$ and
    $\mathsf{C_3}/\mathsf{C_2}$ to $\mathsf{C_5}$.
    Let $c_3$ be a codeword of $\mathsf{C_3}$. We have then
    $c_3=\pi_3(c_3)+c_2$ with $c_2\in \mathsf{C_2}$. We also have
    $c_3=\pi_3(c_3)+\pi_2(c_2)+c_1$ with $c_1\in \mathsf{C_1}$.
    Then the code
    $\mathsf{C'}=\{ (c_3|c_4|c_5) \text{ }\mathrm{with} \text{ }
    c_3\in\mathsf{C_3}, c_5=\varphi_3(\pi_3(c_3))\text{ }\mathrm{and}
    \text{ } c_4=\varphi_2(\pi_2(c_2)) \}$ is self-orthogonal and
    $\mathrm{d}_\mathrm{min}(\mathsf{C'}^\perp)=\mathrm{min}(d_3,d_4,d_5)$
    where $d_3$, $d_4$ and $d_5$ are respective minimum distances of
    $\mathsf{C_3}^\perp$, $\mathsf{C_4}^\perp$ and $\mathsf{C_5}^\perp$.
  \end{theorem}
  \paragraph{Proof :}
  The proof is similar to that of Theorem~\ref{thx} and is therefore
  skipped.
  $\phantom{gauche} \qedwhite$
  \begin{theorem}[Construction X4]
    Let $\mathsf{C_1}$,  $\mathsf{C_2}$, $\mathsf{C_3}$,
    $\mathsf{C_4}$ be four binary
    self-orthogonal linear codes
    of respective lengths $n_1$, $n_1$, $n_3$, $n_3$ and of respective
    dimensions $k_1$, $k_2$, $k_3$, $k_4$ such that
    $\mathsf{C_1}\subset \mathsf{C_2}$, $\mathsf{C_3}\subset
    \mathsf{C_4}$ and $k_2-k_1=k_4-k_3$. Let $\pi_2$ be
    a linear projection from $\mathsf{C_2}$ to
    $\mathsf{C_2}/\mathsf{C_1}$.
    Let $\varphi$ be an
    isomorphism from $\mathsf{C_2}/\mathsf{C_1}$ to
    $\mathsf{C_4}/\mathsf{C_3}$. Then the code
    $\mathsf{C'}=\{(c_2 | \varphi(\pi_2(c_2))+c_3 )
      \text{ }
      \mathrm{with}\text{ } c_2\in\mathsf{C_2}
      \text{ }
      \mathrm{and}\text{ } c_3\in\mathsf{C_3} \}$ is a self-orthogonal
      code of length $n_1+n_3$ and dimension $k_2+k_3$. We also have
      $\mathrm{d}_\mathrm{min}(\mathsf{C'}^\perp)=\mathrm{min}(d_2,d_4)$
      where $d_2=\mathrm{d}_\mathrm{min}(\mathsf{C_2}^\perp)$ and
      $d_4=\mathrm{d}_\mathrm{min}(\mathsf{C_4}^\perp)$.
  \end{theorem}
  \paragraph{Proof :}
  The proof is similar to that of Theorem~\ref{thx} and is therefore
  skipped.
  $\phantom{gauche} \qedwhite$
  
  Additional X type constructions can be found in~\cite{Tjhai2008} and
  similar constructions of self-orthogonal codes can be obtained. All
  these constructions lengthen the size of codewords. There are also
  constructions which can be used to shorten the codes.
  \begin{theorem}[Construction Y1]
    Let $\mathsf{C}$ be a binary self-orthogonal code of length $n$
    and dimension $k$.  Suppose that
    $\mathrm{d}_\mathrm{min}(\mathsf{C}^\perp)=d'$ and let $w$ a
    codeword of $\mathsf{C}^\perp$ of weight $d'$. Let $\mathsf{C'}$
    be the error correcting code obtained by the codewords $c$ of
    $\mathsf{C}$ such that $c_i=0$ if $w_i=1$ and the deletion of the
    $d'$ corresponding coordinates. Then $\mathsf{C'}$ is a binary
    self-orthogonal code of length $n-d'$ and dimension $k-d'+1$.
  \end{theorem}
  \paragraph{Proof :}
  The proof for the parameters of the code is given
  in~\cite[p. 592]{Sloane} and the proof of self-orthogonality is
  obvious.
  $\phantom{gauche} \qedwhite$
  \begin{example}
    Applying construction Y1 to the binary extended $[24,12,8]$ Golay code, we
    obtain a self-orthogonal $[16,5,8]$ code. Its dual is a
    $[16,11,4]$ code and thus a $[[16,6,4]]$ quantum code can be
    built. This code is optimal in terms of minimum distance for CSS
    codes. There is still the possibility that a $[[16,6,5]]$
    stabilizer code exists but no such code is known at the present
    time.
  \end{example}
  \begin{theorem}[Construction Y4]
    Let $\mathsf{C}$ be a binary self-orthogonal code of length $n$
    and dimension $k$.  Let $u$ and $v$, be two codewords of
    $\mathsf{C}^\perp$ such that $u\neq v$ and $d'=w_H(u\text{ } \mathrm{or} \text{ } v)$ is
    minimal amongst all possible pairs. Let then $\mathsf{C'}$
    obtained from codewords of $\mathsf{C}$ such that $c_i=0$ if $u_i
    \text{ } \mathrm{or} \text{ } v_i=1$ and deleting the corresponding coordinates. Then
    $\mathsf{C'}$ is a binary 
    self-orthogonal code of length $n-d'$ and dimension $k-d'+2$.
  \end{theorem}
  \paragraph{Proof :}
  Let $H$ be a parity check matrix of $\mathsf{C}$. This is a
  $(n-k)\times n$ binary matrix. Without loss of
  generality, we can suppose that $u$ and $v$ are in this matrix as
  lines. Let $H'$ be the matrix obtained from $H$ by deleting the
  columns where $u$ or $v$ has a coordinate equal to $1$. This matrix
  has then $n-d'$ columns. If we delete the all null rows
  corresponding to $u$ and $v$, there are $n-k-2$ remaining rows. Let
  $\mathsf{C'}$ be the error correcting code whose parity check is
  $H'$. Its length is therefore $n-d'$ and its dimension is equal to
  $(n-d')-(n-k-2)=k-d'+2$. Any codeword of $\mathsf{C'}$ can be
  converted to a codeword of $\mathsf{C}$ by inserting null bits at
  the coordinates previously corresponding to non null coordinates of
  $u$ and $v$. Reciprocally, any codeword $c\in \mathsf{C}$ such that
  $c_i=0$ if $u_i\text{ } \mathrm{or} \text{ }  v_i=1$ gives a codeword of $\mathsf{C'}$ when
  corresponding coordinates are deleted. Therefore, both definitions of
  $\mathsf{C'}$ coincide. The self-orthogonality of $\mathsf{C'}$ is
  then an obvious consequence of that of $\mathsf{C}$.
  $\phantom{gauche} \qedwhite$
\section{Decoding procedures for classical quantum codes}\label{sec5}
In this section, we describe a decoding procedure that can be used for
CSS codes if the dual of the self-orthogonal code used for their
constructions possesses an algebraic decoding procedure. Then we
illustrate examples for CSS codes defined from self-orthogonal
Reed-Muller and BCH codes.
\subsection{Decoding procedure}\label{decod}
Let $\mathcal{C}$ be a CSS quantum code defined accordingly to
definition~\ref{CSS}. 
A codeword $\mathfrak{c}$ is then emitted over a Pauli
communication channel and a quantum state
$\mathfrak{r}=E\mathfrak{r}$ is received 
at the output, where $E=[E(1)E(2)...E(n)]$ is a length $n$ Pauli operator.
We have then :
\begin{equation}\nonumber
  \begin{split}
    \mathfrak{c} & =\frac{1}{2^{k_2/2}} \sum_{w\in C_2} \ket{c+w}\\
    \mathfrak{r} & =\frac{1}{2^{k_2/2}} \sum_{w\in C_2} E\ket{c+w}
    \end{split}
  \end{equation}
with $c$ being a codeword of $C_1$. The stabilizer associated with
$\mathcal{C}$ is $\mathcal{S}=\mathcal{S_X}\cup \mathcal{S_Z}$ with
$\mathcal{S}_X= \{S_{X,1}, S_{X,2}, ..., S_{X,k_1} \}$ and
$\mathcal{S}_Z= \{S_{Z,1}, S_{Z,2}, ..., S_{Z,k_2} \}$.

We suppose that the transmission has been eventually
corrupted. If $\mathfrak{r}_i=X\mathfrak{c}_i$, it will be said that
there is a bit-flip error in position $i$ of the received qubit frame. If
$\mathfrak{r}_i=Z\mathfrak{c}_i$, it will be said that there is a
phase error in position $i$.  If
$\mathfrak{r}_i=Y\mathfrak{c}_i$, it will be said that there are a
bit flip and a
phase error in position $i$. Let $t_1=\lfloor \frac{d_1-1}{2} \rfloor$
and $t_2=\lfloor \frac{d_2-1}{2} \rfloor$.

Let $S\in \mathcal{S}$. Using classical
techniques~\cite[\S 10.6]{Nielsen}\cite{DiVicenzo}, it is possible to
determine, in a
non destructive way, if $S$ stabilizes $\mathfrak{r}$, i.e. if
$S(\mathfrak{r})=\mathfrak{r}$.
Now, if $S \bigstar E=0$, we have $SE=ES$ and then:
\begin{equation}\nonumber
  \begin{split}
    S(\mathfrak{r}) &= SE(\mathfrak{c}) = ES(\mathfrak{c}) = E(\mathfrak{c})=\mathfrak{r}.
  \end{split}
\end{equation}
On the contrary if $S \bigstar E=1$, we have $SE=-ES$ and it gives
$S(\mathfrak{r})=-\mathfrak{r}$. Therefore $S \bigstar E=0$ if and
only if $S(\mathfrak{r})=\mathfrak{r}$ and otherwise $S \bigstar E=1$.
Therefore, we can determine $s_X(E)=(S_{X,1}\bigstar E,
S_{X,2}\bigstar E, ..., S_{X,k_1}\bigstar E)$ and $s_Z(E)=(S_{Z,1}\bigstar E,
S_{Z,2}\bigstar E, ..., S_{Z,k_2}\bigstar E)$, which are the two parts
of the syndrome $E$ for the stabilizer $\mathcal{S}=\mathcal{S_X} \cup \mathcal{S_Z}$.

Now let $S\in \mathcal{S_X}$, we have then
{\small
\begin{equation}\nonumber
  \begin{split}
    S &\bigstar E= \sum_{k=1}^n
    E(i) \bigstar S(i) \text{ } [ \mathrm{mod} \text{ } 2]\\
    &=\sum_{k=1}^n
\mathbbm{1}(E(i) \in \{Y,Z \},\text{ } S(i)=X) \text{ }[ \mathrm{mod} \text{ } 2].
  \end{split}
\end{equation}
}
Therefore, if we define the length $n$ binary vector $E_X$, such that
$E_{X,j}=\mathbbm{1}(E(j) \in \{Y,Z \}$
for $1\leq j \leq n$, then for $1\leq i \leq k_1$, we have
$S_{X,i}\bigstar E=\mathmbox{<E_X,G_{1i}>}$. If we define similarly
the length $n$ binary vector $E_Z=\mathbbm{1}(E(j) \in \{X,Y \}$,
$1\leq i \leq k_2$, we have
$S_{Z,i}\bigstar E=\mathmbox{<E_Z,G_{2i}>}$. From the syndrome
information given by $s_X(E)$ and $s_Z(E)$, it is then possible to
correct up to $t_1$ phase errors and $t_2$ bit-flip errors.

Figure~\ref{fig1} summarizes the global process.
\begin{figure}[htbp]
  \begin{center}
  \fbox{
    \begin{minipage}{0.45\textwidth}
      {\bf Input:} $\mathfrak{r}$
      with at most $t_1$ phase errors and $t_2$ bit-flip errors.
      \vspace{4pt}
      \hrule
      \vspace{4pt}
      \begin{itemize}
      \item Set $s_X=(s_{X,1},s_{X,2},...,s_{X,k_1})$ such that:
        \begin{equation}\nonumber
          r_{X,i}= \mathbbm{1}(S_{X,i}(\mathfrak{r})=\mathfrak{r})
        \end{equation}
      \item Set $s_Z=(s_{Z,1},s_{Z,2},...,s_{Z,k_2})$ such that:
        \begin{equation}\nonumber
          r_{Z,i}= \mathbbm{1}(S_{Z,i}(\mathfrak{r})=\mathfrak{r})
        \end{equation}
        \item Decode $r_X$, output $E_Z$.
        \item Decode $r_Z$, output $E_X$.
        \item Set 
          \begin{equation}\nonumber
            E = \prod_{i=1}^n
            X^{E_{X,i}\overline{E_{Z,i}}} Y^{E_{X,i}E_{Z,i}} Z^{\overline{E_{X,i}}E_{Z,i}} 
          \end{equation}
        \item Set $\hat{\mathfrak{c}}=E\mathfrak{r}$.
      \end{itemize}
      \vspace{4pt}
      \hrule
      \vspace{4pt}
             {\bf Output:} $\hat{\mathfrak{c}}$, decoded codeword of $\mathcal{C}$
    \end{minipage}
  }
  \end{center}
  \caption{Decoding procedure for CSS quantum codes\label{fig1}}
\end{figure}
In the case when decoding is made by table lookup on the syndrome
values, splitting the problem in two decoding procedure is already a
great improvement. Indeed, instead of having a table of $4^n$ entries
of length $t_1+t_2$, we have now to consider only two tables with
$2^n$ entries of length $t_1$ for the first and $t_2$ for the
second. But, in the case where $C_1^\perp$ and $C_2^\perp$ have
algebraic decoding procedures, there is no more need of tables. In the
following, we describe some families of CSS codes for which this
strategy succeeds.
\subsection{Quantum Reed-Muller codes}\label{qrm}
Quantum Reed--Muller codes have been exposed in~\cite{St}.
They are constructed from the family of binary Reed-Muller codes.
Binary Reed-Muller codes can be defined recursively using the Plotkin
construction introduced in theorem~\ref{plotk}.
\begin{definition}[Reed-Muller codes]
  Let $r$ and $m$ be integers such that $0\leq r <m$ and $m\geq
  2$. The Reed-Muller code of order $r$ in $m$, denoted $RM(m,r)$ is a binary
  error-correcting code of length $n=2^m$, dimension $\sum_{i=0}^r
  \binom{m}{i}$ and minimum distance $2^{m-r}$. By definition,
  $RM(m,0)$ is the repetition code of length $n$ while $RM(m,m-1)$ is
  the parity code of length $n$. If $0<r<m-1$, then $RM(m,r)$ is the
  code obtained by the construction of theorem~\ref{plotk} with
  $\mathsf{C_1}=RM(m-1,r)$ and $\mathsf{C_2}=RM(m-1,r-1)$.
\end{definition}
We have then:
\begin{theorem}
  If $0\leq r <m$, we have $RM(m,r)^\perp=RM(m,m-r-1)$. Moreover, if
  $0\leq r_1 < r_2 <m$, $RM(m,r_1) \subset RM(m,r_2)$.
\end{theorem}
Therefore, we have:
\begin{theorem}
  If $0\leq r\leq \frac{m-1}{2}$, then the code $RM(m,r)$ is
  self-orthogonal. Moreover if $m$ is odd and $r= \frac{m-1}{2}$,
  $RM(m,r)$ is self-dual.
\end{theorem}
The Reed-Muller codes can then be used to construct quantum codes via
the CSS construction (definition~\ref{CSS}) and these latter codes are
called quantum Reed-Muller codes. The classical Reed-Muller codes can
also be decoded up to half the minimum distance by the Reed majority vote
decoding~\cite{Reed}~\cite[p. 385]{Sloane}. For the $RM(m,r)$ code,
the time complexity of this procedure is $O(n \log^r n)$. Thus quantum
Reed-Muller codes can be decoded by our decoding procedure depicted in 
figure~\ref{fig1}.
\subsection{Quantum BCH codes}\label{bch}
Quantum BCH codes have been presented in~\cite{GB}. They are built upon
the family of binary BCH codes~\cite{Hocq,Bose}\cite[ch. 9]{Sloane}.
\begin{definition}[BCH codes]
  Let $n$ being an odd number such that $n | 2^m -1$. Let $\alpha$ be
  a primitive element in the Galois field $\mathrm{GF}(2^m)$ and let
  $\beta=\alpha^{\frac{2^m-1}{n}}$. Let $b\geq 1$ and $\delta\geq 2$
  and let $g$ be the binary polynomial of minimal degree such that
  $g(\alpha^k)=0$ for all $b\leq k \leq b+\delta-2$. Then if
  $\mathrm{deg}(g)<n$, the polynomial $g$ generates a cyclic code,
  named BCH code, of
  length $n$, dimension $n-\mathrm{deg}(g)$ and minimum distance at
  least $2\delta+1$. 
\end{definition}
As an example, the extended Hamming $[8,4,4]$ self-dual code, used to
build the 9-qubit Steane code in the example~\ref{stea9}, can be obtained
by extending the BCH code of length 7 and minimum distance 3.
BCH codes can be decoded up to half the designed minimum distance by
the Peterson-Gorenstein-Zierler~\cite{Peterson,Gorenstein}
procedure. However, for BCH codes with large minimum distance, the
Berlekamp-Massey procedure~\cite{BM} or the Euclidean remainder
Sugiyama procedure~\cite{Sugiyama} are more practical. The time complexity
of these two latter algorithms is comparable and is quadratic with
respect to the minimum distance~\cite[ch. 9 \S6]{Sloane}. However, the
computation of the syndromes and the Chien search increase the time
decoding cost by $O(\delta n)$. The used criterion for
self-orthogonality is derived from the following theorem given
in~\cite{GBP}.
\begin{theorem}\label{thgbp}
Let $\mathsf{C}$ be a cyclic code of length $n$ whose generator
polynomial is $g$. Let $m$ such that $n|2^m-1$, let $\alpha$ be a
primitive element of $\mathrm{GF}(2^m)$ and let
$\beta=\alpha^{\frac{2^m-1}{n}}$. Let then
\begin{equation}\nonumber
I_{\mathsf{C}}=\{i \text{
      s.t. } 0\leq i \leq n-1 \text{ and } g(\beta^i)=0\}.
\end{equation}
\end{theorem}
  We have
  then:
\begin{equation}\nonumber
  I_{\mathsf{C}^\perp}=\{i \text{
    s.t. } 0\leq i \leq n-1 \text{ and } n-i \not\in I_{\mathsf{C}}\}.
\end{equation}
From this we deduce the following theorem.
\begin{theorem}\label{crit}
Let $\mathsf{C}$ be a cyclic code of length $n$ whose generator
polynomial is $g$. Then $\mathsf{C}$ is self-orthogonal if and only if
we have, for all $0\leq i \leq n-1$:
\begin{equation}\nonumber
  g(\beta^{n-i}) \neq 0 \Rightarrow g(\beta^{i}) = 0.
\end{equation}
\end{theorem}
As we have seen in paragraph~\ref{decod}, our decoding procedure
requires a decoding algorithm for $\mathsf{C}^\perp$. Therefore, this
latter code has to be a BCH code to apply the decoding scheme of
paragraph~\ref{decod}. The search of convenient cases is then
organized as follows. Given a length $n$, we enumerate all possible
cases for $b$ and $\delta$ and the associated polynomials are used to
generate $\mathsf{C}^\perp$. Theorem~\ref{thgbp} is then used to
determine the zeros of the generating polynomial of
$\mathsf{C}$. Finally, the criterion of theorem~\ref{crit} is used to
confirm or deny the self-orthogonality of $\mathsf{C}$.
Extended BCH codes can also be used to generate self-orthogonal codes.
\begin{theorem}
Let $\mathsf{C}$ be a self-orthogonal cyclic code with length $n$ and
dimension $k$ such that
$\mathsf{C}^\perp$ is a BCH code whose minimum distance is $2\delta
+1$. Let $\overline{\mathsf{C}^\perp}$ be the extended
BCH code associated with $\mathsf{C}^\perp$ and
$\mathsf{C'}=\overline{\mathsf{C}^\perp}^\perp$.  Then $\mathsf{C'}$
is a self-orthogonal code.
\end{theorem}
\paragraph{Proof :}
The length of the codes $\mathsf{C}$ and $\mathsf{C}^\perp$ is equal
to $n$ and is an odd number since $\mathsf{C}^\perp$ is a BCH code.
Moreover the minimum distance of $\mathsf{C}^\perp$ is odd.
Therefore the length of the codes $\overline{\mathsf{C}^\perp}$ and
$\mathsf{C'}$ is equal to $n+1$. The dimension of $\mathsf{C}$ is
equal to $k$, so that the dimension of $\mathsf{C}^\perp$ and
$\overline{\mathsf{C}^\perp}$ is equal to $n-k$. Therefore the
dimension of $\mathsf{C'}$ is equal to $k+1$. Let $c$ be a codeword of
$\mathsf{C}$ increased by a null symbol at the parity location of
$\overline{\mathsf{C}^\perp}$. Then $c$ is orthogonal to any codeword
of $\overline{\mathsf{C}^\perp}$, so that $c\in C'$. Since the
dimension of $\mathsf{C}$ is equal to $k$, the set of all these
codewords defines a $k$-dimensional vector space in $\mathsf{C'}$.
Let $\mathbf{1}$
be the length $n+1$ all $1$ word. Any codeword of
$\overline{\mathsf{C}^\perp}$ has an even weight and is therefore
orthogonal to $\mathbf{1}$, so that $\mathbf{1}\in C'$. 
Since the last bit of
$\mathbf{1}$ is equal to $1$, it does not belong to vector space of
the previous paragraph. We have then
an entire description of $\mathsf{C'}$. As a consequence, together
with this vector space, $\mathbf{1}$ increase the dimension to $k+1$
which is exactly the dimension of $\mathsf{C'}$. Therefore the
description of $\mathsf{C'}$ is complete.
The codewords of $\mathsf{C'}$
obtained by concatenation of a null symbol are orthogonal with each
other since $\mathsf{C}$ is self-orthogonal. The codeword $\mathbf{1}$
is orthogonal with itself since $n+1$ is even. The codeword
$\mathbf{1}$ is also orthogonal to the codewords of $\mathsf{C'}$
obtained by
adding a null symbol to codewords of $\mathsf{C}$. Indeed, codewords
of $\mathsf{C}$ have even weights because of self-orthogonality. As a
consequence, $\mathsf{C'}$ is a self-orthogonal code.
$\phantom{gauche} \qedwhite$
\section{Codes from finite projective geometries}\label{sec6}
The following definitions and theorems come the exposure~\cite{Veblen}.
\begin{definition}[Finite projective geometries]
  A set $S$ will be called finite projective geometries is defined by the
  following axioms.
  \begin{itemize}
    \item The set $S$ contains a finite greater than $3$ elements
      called points. It also contains lines which are subsets of $S$
      containing at least $3$ points.
    \item If $A$ and $B$ are distinct points of $S$, there is one
      and only one line containing $A$ and $B$.
    \item Let $A$, $B$ and $C$ three non-collinear points and let be
      the lines $l_1$
      containing $A$ and $B$, and $l_2$ containing $B$ and $C$. Let
      then $D$ and $E$ distinct points belonging respectively to $l_1$
      and $l_2$. Then if the line $l$ containing $D$ and $E$ does not
      contain any of the points $A$, $B$ and $C$, the line $l$
      contains a point $F$ belonging to the line $l_3$ containing $A$
      and $C$.
  \end{itemize}
\end{definition}
The spaces of a finite projective geometry are then defined
inductively. The points of the finite projective geometry are each a
$0$-space. The set of $1$-spaces is the set of the lines of the finite
projective geometry.
Then for $k\geq 2$,  if $A_1,A_2,...,A_k$ distinct points belonging to a
$k-1$-space $\mathcal{H}$ and $A_k$ is a point not belonging to
$\mathcal{H}$, the set of points collinear to $A_k$ and any
element of $\mathcal{H}$ forms a $k$-space of the geometry.
A classical example of finite projective geometry is given by Galois
fields.
\begin{theorem}\label{th19}
  Let $k\geq 2$ and $V(k,p^s)$ be the vectorial space of dimension
  $k+1$ over
  $\mathrm{GF}(p^s)$. In $V(k,p^s)$, two vectors will be said to be
  equivalent if they are proportional up to a non null factor of
  $\mathrm{GF}(p^s)$.
  Let $(u_1,u_2,...,u_{k+1})$ be a non-identically
  null vector of $V(k,p^s)$ quotiented by the equivalence
  relation. Then the set of all vectors non-equivalent and non-identically
  null $(x_1,x_2,...,x_{k+1})$ such that:
  \begin{equation}\nonumber
    u_1x_1+u_2x_2+...,+u_{k+1}x_{k+1}=0
  \end{equation}
  forms a $k-1$-space. The $k-2$-spaces are then defined as
  intersections of $2$ distinct $k-1$-spaces. More generally
  $k-l$-spaces are then defined as 
  intersections of $2$ distinct $k-l+1$-spaces. The set of all spaces
  defines then a finite $k$-dimensional projective geometry denoted
  $\mathrm{PG}(k,p^s)$. In this geometry, any $l$-space contains
  exactly $1+p^s+p^{2s}+...+p^{ls}$ points.
\end{theorem}
\begin{figure}
  \begin{center}
  \includegraphics[scale=0.8]{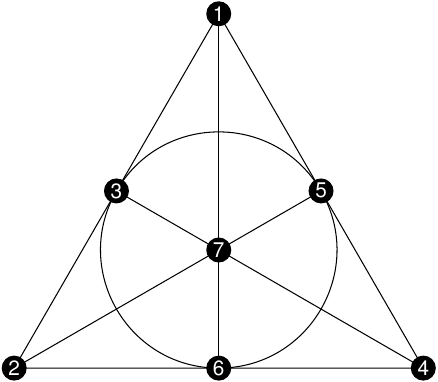}
  \caption{Projective geometry $PG(2,2)$ (Fano triangle)\label{figfano}}
  \end{center}
\end{figure}
In $\mathrm{PG}(k,p^s)$, there are exactly $\frac{p^{(k+1)s}-1}{p^s-1}$
distinct points since the all null vector of $V(k,p^s)$ is excluded
and proportional vectors are equivalent. As an example,
figure~\ref{figfano} depicts the projective geometry
$PG(2,2)$. The $7$ points are represented by the numbered black
circles and the $7$ lines are represented by the $6$ straight segments
plus the internal circle.
The set of lines of
$\mathrm{PG}(k,p^s)$ can then be used to form a binary incidence matrix such
that any two lines of the matrice have both a non null coefficient on
a unique column. This fact has been used in~\cite{Shulin} to
define LDPC codes. This property generalizes to the entire hierarchy
of spaces, since two distinct $l$-spaces intersect in a $l-1$-spaces
and spaces of same rank have same cardinality. The following
definition and results come from~\cite{Rudolf}.
\begin{definition}[Configuration]
  A $(b,v,r,k',\lambda)$-configuration is a system of $b$ sets and $v$
  elements whose incidence $b\times v$ binary matrix has the following
  properties:
  \begin{itemize}
    \item every row has exactly $k'$ $1$'s,
    \item every column has exactly $r$ $1$'s,
    \item any pair of distinct rows have both $1$'s on exactly
      $\lambda$ columns.
  \end{itemize}
\end{definition}
\begin{theorem}
  Let $\mathsf{C}$ be a length $v$ binary error correcting code whose
  generator matrix is a $(b,v,r,k',\lambda)$-configuration. There is a
  majority logic decoding algorithm for $\mathsf{C}^\perp$ which can correct
  up to $\lfloor \frac{r+\lambda-1}{2\lambda} \rfloor$ errors.
\end{theorem}
\paragraph{Proof :}
See~\cite{Ng}.
$\phantom{gauche} \qedwhite$
\begin{theorem}
  For $\mathrm{PG}(k,p^s)$, the incidence matrix related to $l$-spaces
  defines a $(b,v,r,k',\lambda)$-configuration with:
  \begin{equation}
  \begin{split}
    b  = \frac{\prod_{i=0}^l P(i,k)}{\prod_{i=0}^l
      P(i,l)},
    v = P(0,k) ,
    r = \frac{\prod_{i=1}^l P(i,k)}{\prod_{i=1}^l
      P(i,l)},\nonumber \\
    k' = P(0,l),
    \lambda  = \begin{cases}
      1 \text{ if } l=1,\\
      \frac{\prod_{i=2}^l P(i,k)}{\prod_{i=2}^l
      P(i,l)}\text{ if } l>1.
      \end{cases}
  \end{split}
  \end{equation}
  with $P(i,j) = \sum_{m=i}^j p^{ms}$.
\end{theorem}
First, if $k'$ and $\lambda$ are both even, then the code $\mathsf{C}$
defined by the configuration is self-orthogonal. If $k'$ and $\lambda$
are both odd, it is possible to define an self-orthogonal code
$\mathsf{C'}$ of length $v+1$ whose parity matrix will be formed by
the incidence matrix of the configuration plus an additional column
with all the bits set to $1$. This code is still decodable up to
$\lfloor \frac{r+\lambda}{2\lambda} \rfloor$ errors by a two applications
of the majority vote of~\cite{Rudolf}. Suppose that you have
received a noisy codeword of $\mathsf{C'}$. In the first step, you
suppose that the correct $v+1$-th bit is equal to $0$. Then the
decoding of the $v$ other bits can be made directly. In the second
step, suppose that the correct $v+1$-th bit is equal to $1$. Then all
the parities computed by the decoding procedure on the $v$ first bit
will be wrong. It is then enough to add $1$ to obtain the correct
parities and you can apply the procedure with the new parities.
For instance, the binary matrix given in equation~\ref{pg22par} spans the
self-orthogonal code associated with $PG(2,2)$, accordingly to the
numbering of figure~\ref{figfano}, the last column corresponding to
the additional bit.
\begin{equation}\label{pg22par}
  \begin{pmatrix}
    1 & 1 & 1 & 0 & 0 & 0 & 0 & 1\\
    1 & 0 & 0 & 1 & 1 & 0 & 0 & 1\\
    1 & 0 & 0 & 0 & 0 & 1 & 1 & 1\\
    0 & 1 & 0 & 1 & 0 & 1 & 0 & 1\\
    0 & 1 & 0 & 0 & 1 & 0 & 1 & 1\\
    0 & 0 & 1 & 1 & 0 & 0 & 1 & 1\\
    0 & 0 & 1 & 0 & 1 & 1 & 0 & 1
  \end{pmatrix}
\end{equation}
By enumeration of codewords, the minimum distance of the code
generated by this matrix is equal to $4$. Therefore the generated code
is necessarily the self-dual $[8,4,4]$ extended Hamming code. The
generated quantum code is then a $[[8,0,4]]$ quantum code.
For the two other possible cases of parities of $k'$ and $\lambda$, it is still
possible to define parity matrices of self-orthogonal codes, for
instance by the concatenation with a side matrix. If $k'$ is
odd and $\lambda$ is even, you can construct a $b\times (b+v)$ matrix whose
$v$ first columns correspond to the incidence matrix of the
configuration and the last $b$ columns correspond to an identity
matrix of size $b\times b$. If $k'$ is
even and $\lambda$ is odd, in some cases, you can use the same type of
construction where the
rightmost $b\times b$ matrix is such that any two lines have a $1$ in
common and the Hamming weight of any line is odd. It can be done by
using the incidence matrix of a projective geometry if $b$ is an
acceptable value i.e. $b=P(0,s')$ with for some $s'$ and
$k'$. However, in the two
preceding cases, the defined code cannot be decoded by majority
decoding. Thus, we will not consider these situations.
\section{Examples and discussions}\label{sec7}
The enumeration of all codes that could be obtained from the
Reed-Muller, BCH and projective geometry codes have been made up to
length $128$. For the case of Reed-Muller codes, $6$ quantum
error-correcting codes have been obtained. Their parameters are the
following: $[[16,6,4]]$, $[[32,20,4]]$, $[[64,50,4]]$, $[[64,20,8]]$,
$[[128,112,4]]$, $[[128,70,8]]$. For the case of BCH codes, $26$ codes
were obtained and $26$ additional codes can be obtained by adding a
parity bit. The list of these codes is given in Table~\ref{tabbch}.

For codes defined by projective geometries and below length 128, the
possible cases are $PG(2,p^s)$ with $p^s\in \{2,3,4,5,7,9\}$,
$PG(3,p^s)$ with $p^s\in \{2,3,4\}$, $PG(4,p^s)$ with $p^s\in
\{2,3\}$, $PG(5,2)$ and $PG(6,2)$. As seen previously the incidence
matrices associated with the corresponding configurations have a
constant value $\lambda$ for the scalar product of different columns.
However, in order to check the self-orthogonality of the codes, the
scalar product of any pair of rows must at least have the same
parity. And this property does not hold for all the configurations.
For instance, if we consider the projective geometry $PG(4,2)$, we
find that it contains $31$ points. Then using theorem~\ref{th19}, we
can build the entire hierarchy of its spaces. It is found that it
contains $31$ 3-spaces each containing $15$ points, $155$ 2-spaces with
$7$ points each, $155$ 1-spaces with
$3$ points each and $31$ 0-spaces corresponding to the points of the
geometry. The row scalar product of the configuration related to
3-spaces is constant and equal to $7$. For the 2-spaces, the row
scalar product is either equal to $3$ or $1$. Finally for the
1-spaces, the row scalar product is either equal to $1$ or
$0$. Therefore, for 3-spaces and 2-spaces, since the Hamming weight of
each row and the row scalar products are always odd, it is possible to
define a self-orthogonal code by adding an all "1" column. In the case
of $1$-spaces, it is not possible since row scalar products are
either odd or even.

After having checked all the possible cases, it was found that the
projective geometries built with Galois fields of odd characteristic
are not convenient. Up to length 128, the geometries giving
self-orthogonal codes are the following: $1$-spaces of $PG(2,2)$,
$2$-spaces of $PG(3,2)$,  $2$-spaces and $3$-spaces of $PG(4,2)$,
$3$-spaces and $4$-spaces of $PG(5,2)$,  $3$-spaces, $4$-spaces and
$5$-spaces of 
$PG(6,2)$,  $1$-spaces of $PG(2,4)$,  $2$-spaces of $PG(3,4)$ and
$1$-spaces of $PG(3,8)$. Left aside the case of the code built with
$3$-spaces of $PG(6,2)$, the weight spectrum of the codes obtained by
using configurations as generator matrices have been computed. Then,
the weight spectrum of their respective duals was computed by the
Mac-Williams theorem~\cite[p. 128]{Sloane}. Therefore the minimum
distance of all these codes and their duals were obtained.

The
remaining case defines a $[128,64]$ self-dual code. The weight of the
rows of the matrix associated to this code is $16$ and thus is an
upper bound for the minimum distance. Moreover, it implies that this
code is doubly-even and thus any codeword of this code has a Hamming
weight divisible by $4$. Therefore, it is enough to check whether this
code has codewords of weight $4$, $8$ and $12$. A gaussian elimination
was performed on the associated matrix. The obtained matrix acts as a
parity and generator matrix for the code. The $128$ columns of this
matrix can be split in two sets: the set of columns that have been
used as pivots during gaussian elimination and the others. Therefore,
if codewords of weight at most $12$ exist, then necessarily they have
a Hamming weight of at most $6$ either on the set of pivoting columns
or on the set of non-pivoting columns. Thus, we first enumerated all
patterns of at most $6$ pivoting columns and generated the entire
codeword by using the matrix as a generator matrix of the code. Then,
we enumerated all
patterns of at most $6$ non-pivoting columns and generated the entire
codeword by using the matrix as a parity matrix of the code. No
codeword of weight below $16$ was obtained. Therefore the $[128,64]$
self dual code that was obtained has a minimum distance of $16$. This
code is not optimal in terms of minimum distance, since self-dual
$[128,64,20]$ codes are known to exist and at least one $[128,64,22]$
non self-dual code is known. 

\setlength{\tabcolsep}{3pt}
\begin{table}[htbp]
  \begin{center}
    {\scriptsize
      \begin{tabular}{|c|c|}
    \hline
    Code& Generator polynomial generator of BCH code\\
    parameters & (hexadecimal value for integer argument 2)\\
    \hline
    $[[15,7,3]]$ & $\mathrm{0x9AF}$\\
    \hline
    $[[21,9,3]]$ & $\mathrm{0xA4CB}$\\
    \hline
    $[[21,3,5]]$ & $\mathrm{0x1A8F}$\\
    \hline
    $[[31,1,7]]$ & $\mathrm{0x147BF}$\\
    \hline
    $[[31,11,5]]$ & $\mathrm{0x32E8AB}$\\
    \hline
    $[[31,21,3]]$ & $\mathrm{0x6A45F67}$\\
    \hline
    $[[45,13,5]]$ & $\mathrm{0x3A23AD59}$\\
    \hline
    $[[51,35,3]]$ & $\mathrm{0xE326E7B34B1}$\\
    \hline
    $[[55,15,4]]$ & $\mathrm{0xDDD946DFD}$\\
    \hline
    $[[63,51,3]]$ & $\mathrm{0x3F566ED27179461}$\\
    \hline
    $[[63,39,5]]$ & $\mathrm{0xA35C93F631679}$\\
    \hline
    $[[63,27,7]]$ & $\mathrm{0x3320C9F34AF3}$\\
    \hline
    $[[85,69,3]]$ & $\mathrm{0x35ABEA2C24A198F4BB4D}$\\
    \hline
    $[[85,53,5]]$ & $\mathrm{0x3FECD96C8FA9F07243}$\\
    \hline
    $[[89,23,9]]$ & $\mathrm{0x1764DDCBD3B8989}$\\
    \hline
    $[[93,73,3]]$ & $\mathrm{0xEC77E31E49181E3F23EFB}$\\
    \hline
    $[[93,63,5]]$ & $\mathrm{0x703365A734791C2C4EAF}$\\
    \hline
    $[[93,43,7]]$ & $\mathrm{0x1A97E0808F8470F23D}$\\
    \hline
    $[[93,13,11]]$ & $\mathrm{0x3E3E4297282E6B}$\\
    \hline
    $[[127,113,3]]$ & $\mathrm{0x1BE0B087462729A5EBB8F32455B3FB5}$\\
    \hline
    $[[127,99,5]]$ & $\mathrm{0x3190488E5B884A8F2CBF766953B65}$\\
    \hline
    $[[127,85,7]]$ & $\mathrm{0x7B58F033D746D85D06A9F911B4B}$\\
    \hline
    $[[127,71,9]]$ & $\mathrm{0xE2053619F3BBDFFAD8BB92E3F}$\\
    \hline
    $[[127,57,11]]$ & $\mathrm{0x1363666EFD9347B31283796F}$\\
    \hline
    $[[127,43,13]]$ & $\mathrm{0x2612A3178A1AD1832FE6A5}$\\
    \hline
    $[[127,29,15]]$ & $\mathrm{0x73DFA983C0D3A089566B}$\\
    \hline
      \end{tabular}
      }
  \end{center}
\vspace{4pt}
\caption{CSS quantum error correcting obtained from BCH
  codes.\label{tabbch}}
\end{table}

\begin{table}[htbp]
  \begin{center}
    {\scriptsize
      \begin{tabular}{|c|c|c|c|c|c|c|c|}
        \hline
        Geometry & Spaces & $n$ & $k$ & $d$ & $d^\perp$ & $\mathcal{C}$ & $t$\\
        \hline
        $PG(2,2)$ & $1$-sp. & $8$ & $4$ & $4$ & $4$ & $[[8,0,4]]$ & $1$ \\
        \hline
        $PG(3,2)$ & $2$-sp. & $16$ & $5$ & $8$ & $4$ & $[[16,6,4]]$ & $1$ \\
        \hline
        $PG(4,2)$ & $2$-sp. & $32$ & $16$ & $8$ & $8$ & $[[32,0,8]]$ & $3$\\
        \hline
        $PG(4,2)$ & $3$-sp. & $32$ & $6$ & $16$ & $4$ & $[[32,20,4]]$ & $1$ \\
        \hline
        $PG(5,2)$ & $3$-sp. & $64$ & $22$ & $16$ & $8$ & $[[64,20,8]]$& $2$ \\
        \hline
        $PG(5,2)$ & $4$-sp. & $64$ & $7$ & $32$ & $4$ & $[[64,50,4]]$& $1$ \\
        \hline
        $PG(6,2)$ & $3$-sp. & $128$ & $64$ & $16$ & $16$ & $[[128,0,16]]$& $5$  \\
        \hline
        $PG(6,2)$ & $4$-sp. & $128$ & $29$ & $32$ & $8$ & $[[128,70,8]]$& $2$  \\
        \hline
        $PG(6,2)$ & $5$-sp. & $128$ & $8$ & $64$ & $4$ & $[[128,112,4]]$& $1$  \\
        \hline
        $PG(2,4)$ & $1$-sp. & $22$ & $10$ & $6$ & $6$ & $[[22,2,6]]$ & $2$ \\
        \hline
        $PG(3,4)$ & $2$-sp. & $86$ & $17$ & $22$ & $6$ & $[[86,52,6]]$ & $2$ \\
        \hline
        $PG(3,8)$ & $1$-sp. & $74$ & $28$ & $10$ & $10$ & $[[74,18,10]]$ & $4$ \\
        \hline
      \end{tabular}
      }
  \end{center}
\vspace{4pt}
\caption{CSS quantum error correcting obtained from projective
  geometry codes ($n$ is the length of the codes,
  $k=\mathrm{dim}(\mathsf{C})$, $d=\mathrm{d_{min}}(\mathsf{C})$,
  $d^\perp=\mathrm{d_{min}}(\mathsf{C^\perp})$, $t=\lfloor\frac{r+\lambda-1}{2\lambda} \rfloor$).\label{tabpg}}
\end{table}
\setlength{\tabcolsep}{6pt}

The $12$ self-orthogonal codes obtained from projective
geometry codes up
to length $128$ are gathered in the table~\ref{tabpg}. Each line of this
table begins by the projective geometry and the rank of its spaces
which are used to define the configuration. The code $\mathsf{C}$
generated by the configuration has parameters $[n,k,d]$ and is self
orthogonal. Using Rudolph's procedure and the configuration, it is
then possible to
decode the code $\mathsf{C}^\perp$ with parameters
$[n,n-k,d]$. Therefore, it is also possible by the procedure depicted
  in figure~\ref{fig1} to decode the CSS quantum code $\mathcal{C}$
  with parameters $[[n,n-k,d]]$. The last column gives the parameter
  $t$ of the configuration which is the maximum number of errors that
  can be decoded by Rudolph's procedure.
We can note that, amongst the generated codes,
$3$ of them are in fact self-dual and define $0$-dimensional quantum
codes. We can also remark that codes obtained from projective
geometries over $GF(2)$ have minimum distances equal to a power of
$2$. This is probably a general property. In terms of minimum distance,
according to~\cite{Grassl:codetables}, the codes of length $16$, $32$ 
are optimal as well as the codes of parameters $[64,7,32]$,
$[64,57,4]$ and $[22,12,6]$. As for the codes of parameters
$[64,42,8]$, $[86,69,6]$ and $[74,46,10]$, these codes are not optimal
in regard of theoretical bounds but have the same minimum distance
as the best-known codes.
The last column of the table gives,
for each dual code, the
maximum of errors that can be corrected by Rudolph's majority vote
procedure. Therefore it also represents the power of algebraic
decoding procedure codes. It is then noteworthy that each of the
obtained codes, except the $[128,64,16]$ self-dual code and the
$[64,42,8]$ self-orthogonal code, can be decoded
up to its true minimum distance.
\section*{Conclusion}
In this article, new quantum error correcting codes have been
proposed. A procedure of decoding for CSS quantum codes built upon
classical codes with an algebraic decoding procedure have been
proposed. Techniques to build new codes from old ones have also been
described. A certain number of future axes of research can be
investigated. For instance, it could be possible to explore other
families of codes
with decoding procedures. Another perspective is to generalize this
work to the general family of stabilizer codes. It can effectively be
shown that for a given set of parameters stabilizer codes can have
better minimum distance than the CSS codes.

\end{document}